# Atomic-signal-based zero field finding technique for unshielded laser-pumped atomic magnetometer

Haifeng Dong, Hongbo Lin, Xinbin Tang

*School of Instrumentation Science and Opto-electronics Engineering, Beihang University, Science and Technology on Inertial Laboratory, Fundamental Science on Novel Inertial Instrument & Navigation System Technology Laboratory, 37#, Xuyuan Road, Beijing, 100191, China*

We described a novel technique that can find the zero-field for unshielded laser-pumped atomic magnetometer using atomic signal itself. By comparing light density of pump beam after atomic vapor cell, it is decided which direction to move the compensation magnetic field and whether to increase or decrease the converging step length. The zero-field is found in less than 18s and the step length after converging is smaller than 10nT, 10nT and 40nT for x, y and z axes, respectively, limited by 50Hz noise in the lab environment.

High sensitivity magnetometer measuring the magnetic field vectors in an unshielded environment has a wide range of applications from monitoring the earth's magnetic field, detecting magnetic anomalies to deep space exploration [1-3]. Although SQUID magnetometer or gradiometer were reported to be used for such high sensitivity applications[1, 3], they require cryogenic cooling, which increases the volume and cost of the measurement system [4]. In recent years, atomic magnetometer has emerged as a promising non-cryogenic, low-cost alternative to SQUID magnetometer. Atomic magnetometer can achieve sub-fT sensitivity which opens up new possibilities for ultrasensitive magnetometry [5-7]. However, atomic magnetometer has a very narrow dynamic range and usually works under shielded environment such as magnetoencephalography [8, 9], material characterization [10] etc.. In order to take advantage of the inherent sensitivity of atomic magnetometer for unshielded applications, S. J. Seltzer and M.V. Romalis proposed the method of cross modulation and used magnetic feedback with Helmholtz coils to ensure that atomic vapor cell work under zero fields which for the first time demonstrated the feasibility of atomic magnetometer in the measurement of the unshielded field [11]. Because the earth field amplitude is much larger than the dynamic range of atomic magnetometer, zeroing the field using only atomic signal is extremely difficult and inefficient. One method is to use a different sensor, such as a fluxgate, and then turn on feedback from atomic magnetometer once the field along all three directions is sufficiently small [11, 12]. In this paper, a novel technique to zero the field by atomic signal itself using digital controlling system and smart algorithm is demonstrated. Besides the direct purpose of zeroing field, this method can also be used to detect DC or low frequency magnetic field.

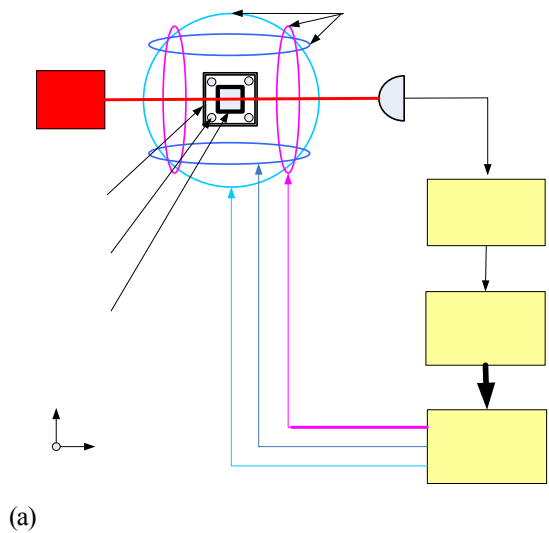

(a)

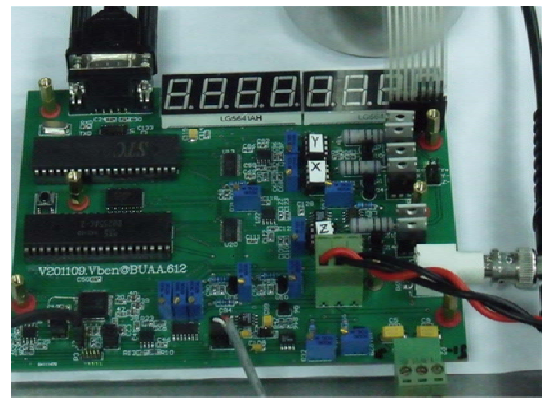

(b)

Fig. 1. The measurement setup of atomic magnetometer self-compensation system (a) and the sampling circuit and the controller (b)

The measurement setup is illustrated in Fig. 1(a). The vapor cell containing Cs atoms, one amagat of buffer gas and 30 Torr of $N_2$ for quenching, is heated to 90°C

~ 120°C by four heaters with high frequency current of 100mA@30kHz. The cell is a cylinder of Φ20×30mm. A DFB laser circularly polarized is tuned to D1 resonance at 894nm to pump the atoms in the vapor cell. Three dimensional coils are used to compensate the earth magnetic field. The current supply is controlled by the convergence algorithm which gets atomic spin signal from the pump laser behind the atomic vapor cell by photodetector (PD) and data acquisition system. Fig.1 (b) shows the hardware in which the convergence algorithm is executed.

The convergence algorithm is based on the stable solution of Bloch equation of atomic spin polarization, the detail of which can be found in [12-15]. The algorithm flow process chart is shown in Fig. 2.

Firstly, the initial values of Bx0, By0 and Bz0 are set by roughly approximating the earth magnetic field in the lab environment. Then the initial values of deltaBx, deltaBy and deltaBz are set to 2000nT, respectively. After that, Bx0, By0 and Bz0 are applied by the current supply in Fig. 1 (a). The light density of pump beam after the atomic vapor cell is detected and sampled as Ph0. Next, Bx0-deltaBx, By0 and Bz0 are applied and the light density is sampled as Ph_minus. Then Bx0+deltaBx, By0 and Bz0 are applied and the light density is sampled as Ph_plus. After that the value of Ph0, Ph_plus and Ph_minus are compared. Given that the total magnetic field along x-axis after compensation is Bx_t, Bx_minus_t and Bx_plus_t corresponding to Bx0, Bx0-deltaBx and Bx+deltaBx, respectively, if Ph_plus<Ph_0 and Ph_minus<Ph_0, it means that the zero magnetic field along x-axis is between Bx_minus_t and Bx_plus_t, then deltaBx is set to deltaBx/2. If Ph_plus>Ph0>Ph_minus or Ph_plus<Ph0<Ph_minus, it means that the zero magnetic field is not in the range between Bx_minus_t and Bx_plus_t. In the first case, Bx0 is set to be Bx0+deltaBx, and in the second case Bx0 is set to be Bx0-deltaBx in order to move the range closer to zero fields. The case of Ph_plus<Ph_0 and Ph_minus<Ph_0 will not happen until the resolution of the measurement is larger than the output difference caused by deltaBx, or the frequency of the measurement is slower than the variation of the environment fields. In this case deltaBx is set to deltaBx×2.

The above process is for x-axis. After that, the similar process is performed for y-axis and z-axis. The cycle is continued until all the three orthogonal magnetic field vectors are compensated to zero. The uncertainty of the compensation is decided by the resolution and bandwidth of the measurement system which is shown by the value of deltaBx, deltaBy and deltaBz after converging.

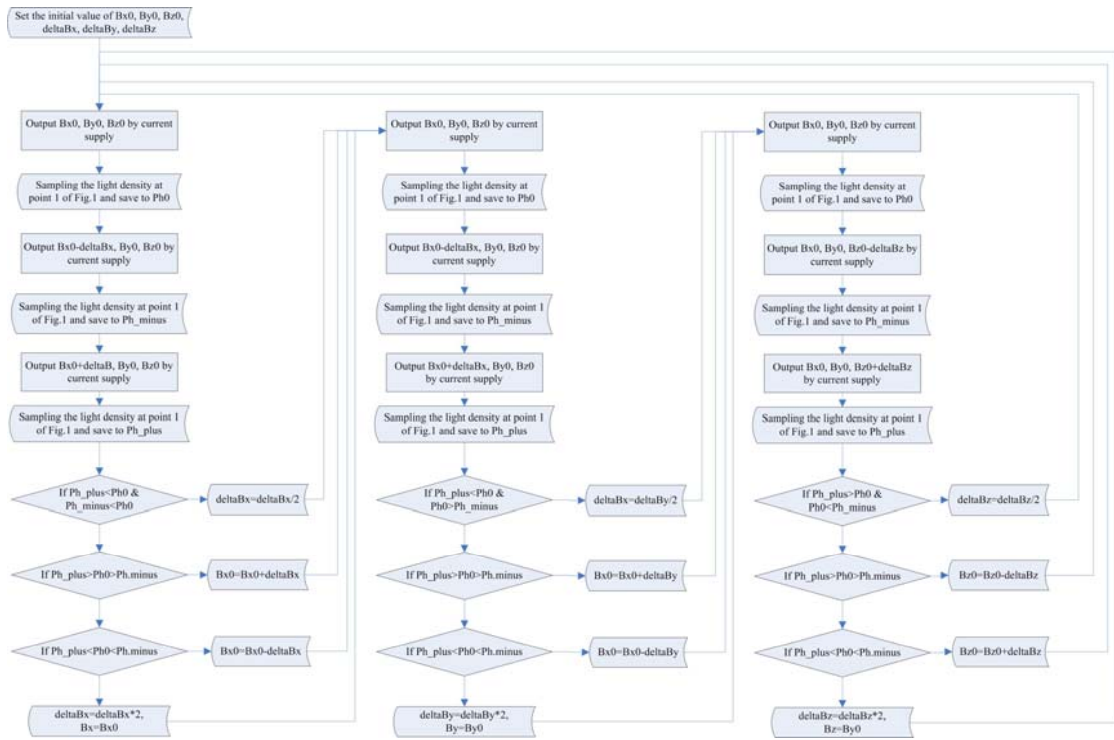

Fig. 2 The convergence algorithm flow process

Fig. 3 is the converging process of the measurement system. To illustrate the progress better, a scanning process is added after the y-axis cycle. The scanning range is from By-1500nT to By+1500nT. From fig. 3, we can see that (1) during the convergence process, the scanning range of the total magnetic field is approaching ±1500nT and the scanning shape is becoming symmetry gradually, in other words, the total magnetic field along y-axis approaches zero gradually; (2) the linewidth of the scanning is reduced because the total magnetic fields are decreased; (3) the 50Hz disturbing can be observed after the fields are compensated close to zero, the amplitude of which is equivalent to about 100~200nT.

DeltaBx, deltaBy and deltaBz after converging are smaller than 10nT, 10nT and 40nT for x, y and z-axes, respectively, which are limited by the 50Hz magnetic disturbing shown in fig. 3. The linewidth of the scanning is about 600nT even when the Cs vapor cell is heated to 120℃which means that the self-exchange relaxation is still the main relaxation mechanism for the Cs atoms. To achieve the self-exchange relaxation free (SERF) regime, one needs to shield the 50Hz noise or increase the bandwidth of the algorithm operation higher than 50Hz. As Bx, By and Bz compensate the environment fields to zero, they can be treated as the measurement results of the fields with opposite sign.

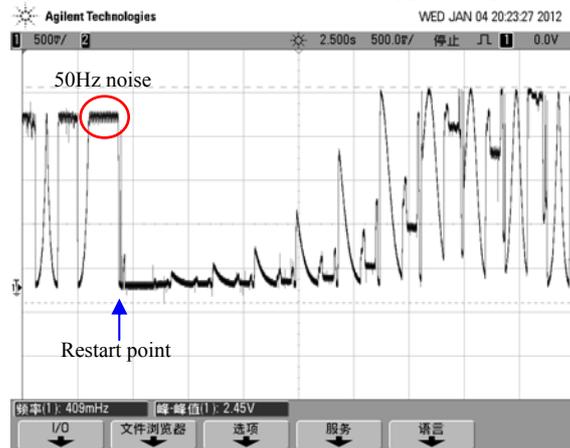

Fig. 3 the converging process of the measurement system

In conclusion, a new method is realized for zeroing magnetic field of unshielded atomic magnetometer. The method is based on direct measurement of the atomic signal instead of using other magnetometer such as fluxgate. The measurement presented here demonstrates that it is possible to zero the field using atomic signal itself and realize self-test of precision for atomic magnetometer using digital controller and smart algorithm. Improved algorithm and higher speed controlling circuit in the future may help to realize SERF regime in the earth environment and enhance the sensitivity of unshielded magnetic measurement.

This work was supported by NSF of China under Grant No. 61074171, 60736025 and Beijing NSF under Grant No. 3122025. The authors also would like to thank Jie Qin, Tao Wang and other Ph.D candidates in atomic device group of BUAA for their beneficial discussions. Haifeng Dong's email address is: donghaifeng.shanxi@yahoo.cn